\documentclass[twocolumn,prl, showpacs]{revtex4}
\usepackage{graphics,graphicx}
\begin{document}
\title{Quasi-local critical nature of cooperative paramagnetic fluctuations in CaRuO$_{3}$ metal}
\author{J.~Gunasekera$^{1}$}
\author{L. ~Harriger$^{2}$}
\author{T. ~Heitmann$^{3}$}
\author{H.~Knoll$^{1}$}
\author{D. K.~Singh$^{1,*}$}
\affiliation{$^{1}$Department of Physics and Astronomy, University of Missouri, Columbia, MO 65211}
\affiliation{$^{2}$NIST Center for Neutron Research, Gaithersburg, MD, 20899}
\affiliation{$^{3}$University of Missouri Research Reactor, University of Missouri, Columbia, MO 65211}
\affiliation{$^{*}$email: singhdk@missouri.edu}

\begin{abstract}

We report new observation of cooperative paramagnetic fluctuations of Ru$^{4+}$ spins that coexist with the non-Fermi liquid state in CaRuO$_{3}$ perovskite below T $\simeq$ 22 K. Detailed electrical, magnetic and neutron scattering measurements reveal that the Ru$^{4+}$ ions reside in magnetic field independent random domains with dynamic properties that are reminiscent of the cooperative paramagnetic fluctuations. The linear ($E/T$) scaling of the dynamic susceptibilities and divergence of the mean relaxation time as $T$$\rightarrow$~0 K suggest quasi-local critical nature of the spin fluctuations. We argue that the non-Fermi liquid behavior arises due to the quantum critical nature of the cooperative paramagnetic fluctuations in CaRuO$_{3}$.
 
\end{abstract}

\pacs{71.10.Hf, 75.40.-s, 78.70.Nx, 71.27.+a} \maketitle

The presence or absence of the Fermi liquid behavior in perovskite materials holds strong promises in understanding the exotic electronic properties that often accompany at low temperature, such as unconventional superconductivity and the quantum criticality.\cite{Lohneysen,Broun,Cooper} In addition to the fundamental understanding, ruthenates perovskite of the form $A$RuO$_{3}$, where $A$ = Sr or Ca, have possible implications in the technologically important spintronics devices, especially in the thin film form where magnetic properties can easily be manipulated for desired applications as the spin-polarized tunnel junctions.\cite{Takahasi,Tripathi} While SrRuO$_{3}$ is ferromagnetic ($T$$_{c}$$\simeq$~160 K), the magnetic nature of CaRuO$_{3}$ is still a subject of debate.\cite{Kikugawa,Longo,Cao2,Cava,Mazin,Cao1,Khalifah,Tripathi,Felner} Both compounds, however, exhibit interesting non-Fermi liquid (NFL) behavior at low temperature ($T$$\leq$40 K).\cite{Kostic,Lee, Cao1,Klein} CaRuO$_{3}$, in fact, undergoes a cross-over transition at $T$~$\simeq$~30 K, which separates two NFL states with distinct power law exponents of electrical resistivity.\cite{Klein} In a more recent work, Schneider et al. have claimed the observation of the quantum oscillations, implying a Fermi-liquid ground state, in a thin film of CaRuO$_{3}$ at very low temperature $T$$\leq$~1.5 K; thus, challenging the present understanding of this material.\cite{Schneider} The unusual combination of the absence of magnetic order and the anomalous non-Fermi liquid properties makes CaRuO$_{3}$ an archetypal perovskite for the exploration of the quantum magnetism.\cite{Georges}

In this report, we address the important question of the dynamic magnetism and its relation to the non-Fermi liquid behavior in CaRuO$_{3}$ using detailed electrical, magnetic and neutron scattering measurements on high quality polycrystalline sample of CaRuO$_{3}$. For the first time, we show that the underlying magnetism is depicted by the quantum mechanical fluctuations of Ru$^{4+}$ spins, confined to field-independent random domains that form a cooperative paramagnetic state at low temperature. The dynamic structure factor, which increases significantly below $T$$\leq$22 K, manifests a linear ($E/T$) scaling, implying the Curie-Weiss type fluctuations with temperature as the most relevant parameter. Moreover, the linear dynamic scaling in conjunction with the divergence of the spin fluctuations mean relaxation time as $T$$\rightarrow$~0 K suggest the existence of a $^{'}$quasi-local$^{'}$ critical behavior in the system. Finally, we argue that the quantum spin fluctuations arising due to the dynamic cooperative paramagnetism is at the core of the low temperature ($T$$\leq$30 K) non-Fermi liquid properties in CaRuO$_{3}$.

CaRuO$_{3}$ is a pseudocubic metallic perovskite, which crystallizes in the orthorhombic structure ($Pnma$ crystallographic group) with lattice parameters of $a$ = 5.545 $\AA$, $b$ = 7.673 $\AA$ and $c$ = 5.398 $\AA$.\cite{Cava,Cao1,Kikugawa, Akaogi} The octahedral crystalline electric field splits the fivefold degeneracy of Ru 4$d$$^{4}$ into ($t$$_{2g}$)$^{4}$ and empty $e$$_{g}$, thus resulting in a low-spin electronic configuration.\cite{Mazin,Cao2}. $^{17}$O NMR investigation of the spin dynamics in Sr$_{1-x}$Ca$_{x}$RuO$_{3}$ claimed robust ferromagnetic fluctuations of Ru$^{4+}$ spins for all doping percentages.\cite{Yoshimura} The inevitable introduction of disorder due to the chemical doping complicates such universal analysis.\cite{Capogna} Heat capacity measurement on single crystal CaRuO$_{3}$ has revealed a logarithmic temperature dependence with a reasonably heavy electron characteristics, specific heat coefficient of 73 mJ$/$molK$^{2}$, at low temperature\cite{Cao1,Shepard}. These observations are reminiscent of the local quantum critical phenomenon, usually found in the intermetallic heavy fermion systems.\cite{Schroder,Si} It is long suspected that the non-Fermi liquid behavior in CaRuO$_{3}$ culminates to a quantum critical state as $T$$\rightarrow$0 K, but has never been observed.

\begin{figure}
\centering
\includegraphics[width=8.7 cm]{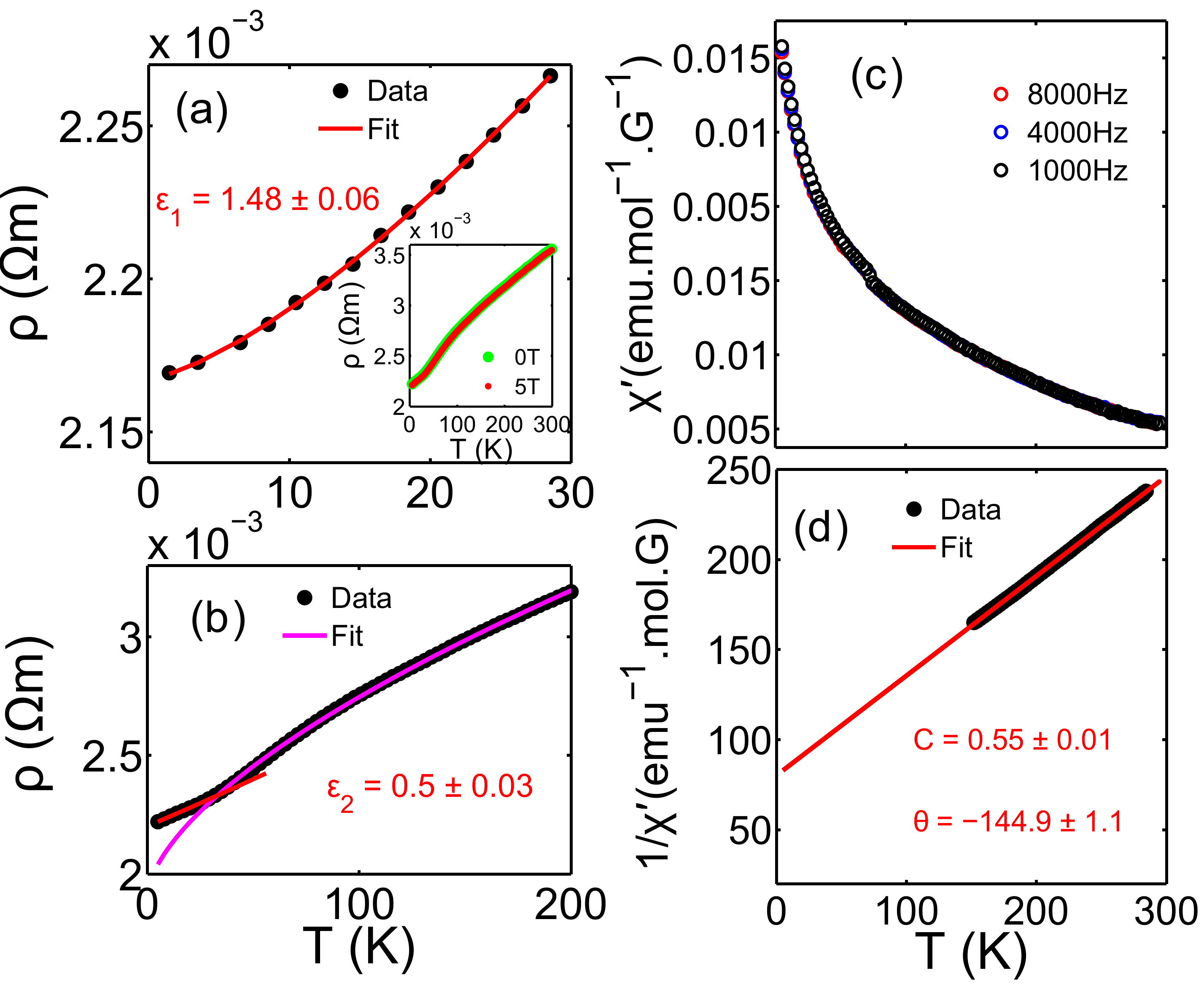} \vspace{-4mm}
\caption{(color online) Electrical resistivity and magnetic measurements of CaRuO$_{3}$ (a) and (b) Electrical resistivity as a function of temperature, depicting two different non-Fermi liquid states in low temperature (below $T$$\simeq$~ 30 K) and high temperature (30 K$\leq$T$\leq$200 K) regimes. The resistivity data is fitted using the power-law equation, $\rho$=$T$$^{\epsilon}$, with fitted value of $\epsilon$ = 1.48(0.06) (Fig. 1a) and 0.5(0.03) (Fig. 1b) in low and high temperature regimes, respectively. Electrical resistivities as a function of temperature at different applied magnetic fields, $H$ = 0 and 5 T, are shown in the inset. No significant field effect on the electrical resistance is observed. (c) Static susceptibilities ($\chi$$^{'}$) at few characteristic frequencies are plotted as a function of temperature in this figure. $\chi$$^{'}$ is found to be independent of the ac frequency. We also observe a small but sharp cusp in $\chi$$^{'}$ at $T$$\simeq$ 80 K. While the origin of this cusp is not yet clear, no magnetic order was detected in the elastic neutron scattering measurements below this temperature. (d) High temperature ($T$$\geq$150 K) inverse static susceptibility data, at very low frequency ($\simeq$ 10 Hz), is fitted using the Curie-Weiss (CW) law. 
} \vspace{-4mm}
\end{figure}

The high purity polycrystalline samples of CaRuO$_{3}$ were synthesized by conventional solid-state reaction method using ultra-pure ingredients of RuO$_{2}$ and CaCO$_{3}$ (see Supplementary Materials for detail). Resulting samples were characterized using powder X-ray diffraction method, confirming the single phase of material (see Fig. S1 of the Supplementary Materials). Four probe technique was employed to measure electrical properties of CaRuO$_{3}$ using a closed-cycle refrigerator cooled 9 T magnet with measurement temperature range of 1.5-300 K. Detailed ac susceptibility measurements were performed using the Physical Properties Measurement System with a temperature range of 2-300 K. Neutron scattering measurements were performed on 4.8 g pristine powder sample of CaRuO$_{3}$ on the spin-polarized triple-axis spectrometer (SPINS) at the NIST Center for Neutron Research and on the thermal triple-axis spectrometer TRIAX at the Missouri University Research Reactor with fixed final neutron energies of 3.7 meV and 14.7 meV, respectively. At these fixed final energies, the spectrometers resolution (FWHM) were determined to be $\simeq$ 0.16 and 0.8 meV, respectively. Measurements on SPINS employed a cold Be-filter followed by a radial collimator and the focused analyzer. Measurements were also performed in applied field with the sample mounted at the end of a 1K stick in a 10 T vertical field magnet. 

The temperature dependence of the electrical resistivity is shown in Fig. 1a and 1b. The experimental data are well described by a power-law expression, $\rho$$\propto$~$T$$^{\epsilon}$. The fitting parameter $\epsilon$ is found to exhibit two different values: $\epsilon_1$ = 1.48(0.06) below $T$ $\simeq$ 30 K and $\epsilon_2$ = 0.5(0.03) in 30 K$\leq$T$\leq$200 K; illustrating two different non-Fermi liquid states in temperature. These results are consistent with the previous observations on high quality specimens of CaRuO$_{3}$.\cite{Klein,Lee,Cao1,Kikugawa} Surprisingly, the resistivity is found to be independent of the magnetic field application (upto 9 T field), as shown in the inset of Fig. 1a. Measurements of the static magnetic susceptibility ($\chi$$^{'}$) reveal two interesting behaviors (Fig. 1b): a small cusp around $T$ $\simeq$ 80 K and the divergence of $\chi$$^{'}$ below $T$ $\simeq$ 22 K. A similar phenomenon (the cusp at $T$$\simeq$~80 K) was previously attributed to the onset of a long range magnetic order.\cite{Felner} However, no signature of any magnetic order is observed in the elastic neutron scattering measurements (see Fig. S2 in the Supplementary Materials). This anomalous observation can be associated to a weak structural transition, which we cannot resolve in the elastic neutron scattering data on the polycrystalline sample. Bulk static susceptibilities, including the cusp at $T$ $\simeq$ 80 K, are also found to be independent of ac measurement frequency, hence rules out the occurrence of the spin freezing or a short-range static order in the system. The divergence in $\chi$$^{'}$ below $T$$\simeq$~22 K is most likely arising due to the quantum critical nature of the cooperative paramagnetic fluctuations, as explained later.

\begin{figure}
\centering
\includegraphics[width=8.9 cm]{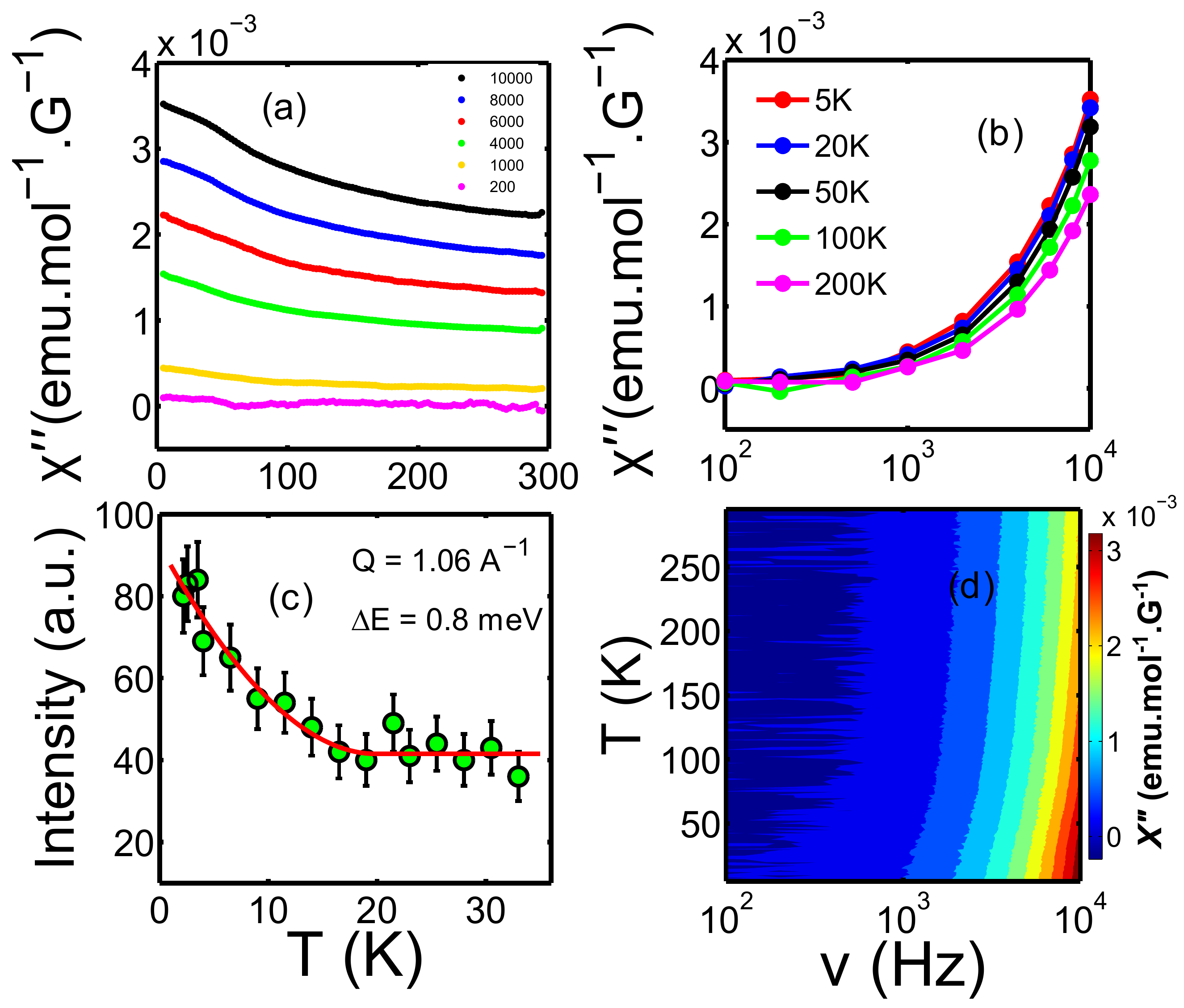} \vspace{-4mm}
\caption{(color online) Dynamic magnetic properties of CaRuO$_{3}$. (a) Dynamic ac susceptibilities ($\chi$$^{"}$) are plotted as a function of temperature at few characteristic frequencies. Clearly, the system exhibits significant dynamic response to the ac frequency (10 Hz - 10$^{4}$ Hz), which becomes stronger below $T$$\simeq$ 25 K. (b) $\chi$$^{"}$ as a function of ac frequency at few different temperatures are shown in this figure. We immediately notice the divergence in $\chi$$^{"}$ at high frequency at any temperature. (c) Inelastic neutron scattering intensity at $Q$ = 1.06 $\AA$$^{-1}$ and at an energy transfer of $\Delta$E = 0.8 meV is plotted as a function of temperature in this figure. The error bar represents one standard deviation. Inelastic intensity, which exhibits significant enhancement below $T$$\simeq$21 K, is described by a power-law equation, yielding $\beta$ = 0.45 (0.017). (d) $T$ vs $\nu$ contour map of $\chi$$^{"}$ sums up the dynamic behavior of magnetic fluctuations as inferred from the ac susceptibility measurements. 
} \vspace{-4mm}
\end{figure}

Fitting of the high temperature (T $\geq$ 150 K) static susceptibility data using the Curie-Weiss law $\chi$$^{'}$ = C/($T$-$\theta$), Fig. 1c, yields $\theta$ = -144.9 K and C = 0.55. Under the mean-field approximation, the Curie-Weiss temperature of -144.9 K implies an antiferromagnetic interaction of strength $J$ $\simeq$ 6.8 meV for the effective moment of 2.56 $\mu$$_{B}$ per ruthenium ion (estimated from the CW constant C).\cite{MFT} Despite a reasonably large exchange constant and a near full moment value in the Curie-Weiss regime, the absence of magnetic order is perplexing. Recent reports, particularly on the thin film of CaRuO$_{3}$, suggest that the system is on the verge of attaining an ordered state. Another possibility is that strong spin fluctuations causes destabilization to a nascent order, thus preventing it from developing a long range magnetic order. We have further investigated the dynamic properties to explore this possibility. 

Plots of the dynamic susceptibilities ($\chi$$^{''}$) as functions of temperature and frequency are shown in Fig. 2a and 2b, respectively. Unlike the static susceptibilities, the dynamic susceptibilities exhibit strong frequency dependence. In an interesting observation, we notice a significant enhancement in $\chi$$^{"}$ as the strength of ac frequency increases from 10 Hz to 10$^{4}$ Hz (Fig. 2a). The dynamic susceptibilities tend to diverge as the measurement temperature is reduced below $T$ $\simeq$ 25 K. Coincidentally, this is also the cross-over temperature separating two non-Fermi liquid states at low temperature.\cite{Klein} The frequency dependence of $\chi$$^{''}$ hints of significant spin fluctuations in CaRuO$_{3}$. The dynamic behavior is also reflected in the inelastic neutron scattering measurement where the dynamic structure factor, depicted by the neutron intensity count, at $Q$ = 1.06 $\AA$$^{-1}$ and at an energy transfer of $\Delta$$E$ = 0.8 meV exhibits a power-law dependence in temperature below $T$ $\simeq$ 21 K, Fig. 2c. The temperature vs. frequency (energy) behavior, describing the dynamic response of Ru$^{4+}$ spins to the external tuning parameters of frequency and temperature, is described in a comprehensive contour plot in Fig. 2d. Clearly, the spin fluctuations get stronger as the temperature is reduced, indicating quantum characteristic of the behavior. 

\begin{figure}
\centering
\includegraphics[width=7.8 cm]{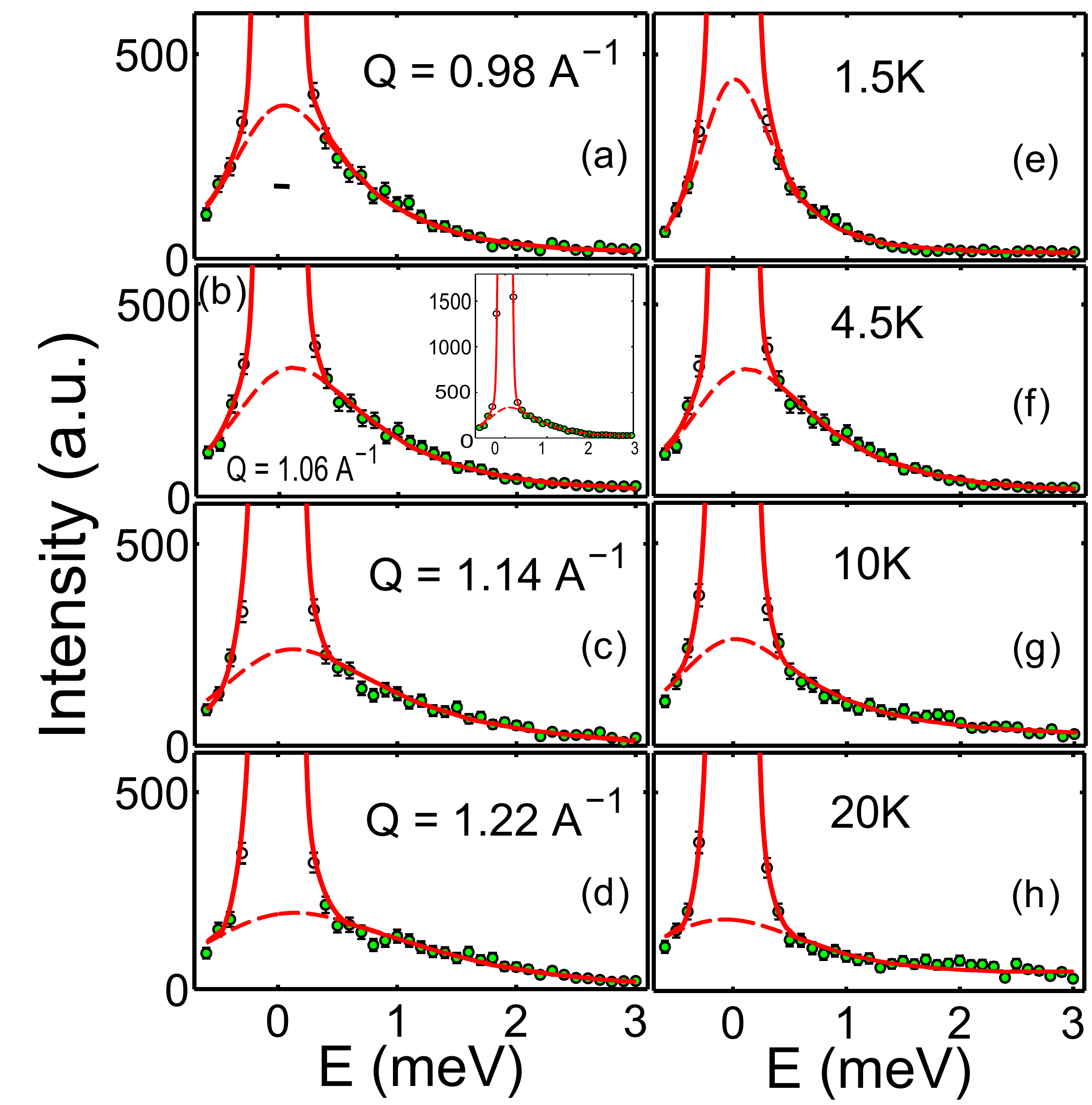} \vspace{-4mm}
\caption{(color online) Inelastic neutron scattering measurements, carried out on SPINS, as a function of energy. (a-d) Characteristic energy scans at constant $Q$ at $T$ = 4.5 K are plotted in these figures. The solid line represents the complete fitting profile of the background corrected experimental data due to Eq. (1), convoluted with the instrument resolution and consists of an elastic line (Gaussian) and a quasi-elastic function multiplied by the detailed balance factor. The quasi-elastic data (dashed lines) are well described by a Lorentzian-squared line-shape, emphasizing the cooperative paramagnetic nature of Ru$^{4+}$ spins fluctuation in the system (see text for detail). As the absolute $Q$ increases, the peaks become broader and weaker, suggesting magnetic nature of fluctuation. (e-h) Energy scans at constant temperatures at $Q$ = 1.06 $\AA$$^{-1}$ are plotted. The inelastic spectra becomes stronger as the measurement temperature is reduced. The error bar in all figures represents one standard deviation.
} \vspace{-4mm}
\end{figure}

In order to elucidate the quantum mechanical nature of the spin fluctuations in CaRuO$_{3}$, detailed inelastic neutron scattering measurements were performed on SPINS as a function of energy with fixed final energy of 3.7 meV, yielding a resolution of $\simeq$ 0.16 meV. Inelastic data were collected at various fixed $Q$-values between 0.7 and 1.5 $\AA$$^{-1}$ and at various temperatures between $T$ = 1 K and 120 K. Background corrected characteristic energy scans at selected $Q$ values at $T$ = 4.5 K and at selected temperatures at Q = 1.06 $\AA$$^{-1}$ are plotted in Fig. 3a-d and Fig. 3e-h, respectively. Magnetic nature of the quasi-elastic spectra is ascribed to the observed weakening of the peak intensity at higher $Q$ and the weak spectral count above the background at high temperatures. Inelastic measurements were also performed in applied magnetic field but no significant or observable difference compared to the zero field data was detected at any temperature (illustrated in Fig. S3 of the Supplementary Materials).

Inelastic neutron scattering data were fitted using various relevant formulations to understand the dynamic properties of Ru$^{4+}$ spins. The experimental data are found to be best described by a cooperative paramagnetic fluctuation model, involving higher orders of Lorentzian function:\cite{Shirane}
\begin{eqnarray*}
{I(Q, E)}&=& {exp(-(\frac{E}{\Gamma_0})^{2})}+({\frac{\Gamma_1}{{E^{2}}+{\Gamma_1}^{2}}})^{\nu}\frac{E/k_BT}{1-{e}^{-E/k_BT}}
\end{eqnarray*},
where $\Gamma_0$ and $\Gamma_1$ are fitting parameters, related to the actual line-widths of the elastic and the quasi-elastic features, respectively. The solid line in Fig. 3 manifests the complete fitting profile of the experimental data due to Eq. (1), convoluted with the instrument resolution and consists of an elastic line (Gaussian) and a quasi-elastic function (higher order Lorentzian function) multiplied by the detailed balance factor.The best fit is obtained for $\nu$ = 2, giving a Lorentzian-squared lineshape of the quasi-elastic data. The width of the elastic line, FWHM of the Gaussian profile, remains resolution limited ($\simeq$0.16 meV) at all temperatures. The Lorentzian-squared lineshape, arguably, manifests random fields arising from the impurities or the presence of magnetic domains or clusters in a system, for example a spin glass or geometrically frustrated magnet.\cite{McDougall} The high purity of the sample rules out impurity as the underlying reason. The formation of small ferromagnetic domains, behaving as dynamic paramagnetic clusters, of Ru$^{4+}$ spins is a possible explanation for this unusual observation. At the same time, however, both elastic and inelastic measurements remain unaffected to the magnetic field application of strength up to 10 T (see Fig. S2b and Fig. S3 in the Supplementary Materials). This is in stark contrast to the previous understanding that the system is $^{''}$on the verge of the ferromagnetic instability$^{''}$.\cite{Cava,Cao1,Mazin} Rather, we argue that the magnetic ground state in CaRuO$_{3}$ consists of field-independent domains that are strongly fluctuating as cooperative paramagnets. The cooperative paramagnetic fluctuations was previously used to describe a spin liquid type ground state in Tb$_{2}$Ti$_{2}$O$_{7}$.\cite{Gardner} 

\begin{figure}
\centering
\includegraphics[width=8.5 cm]{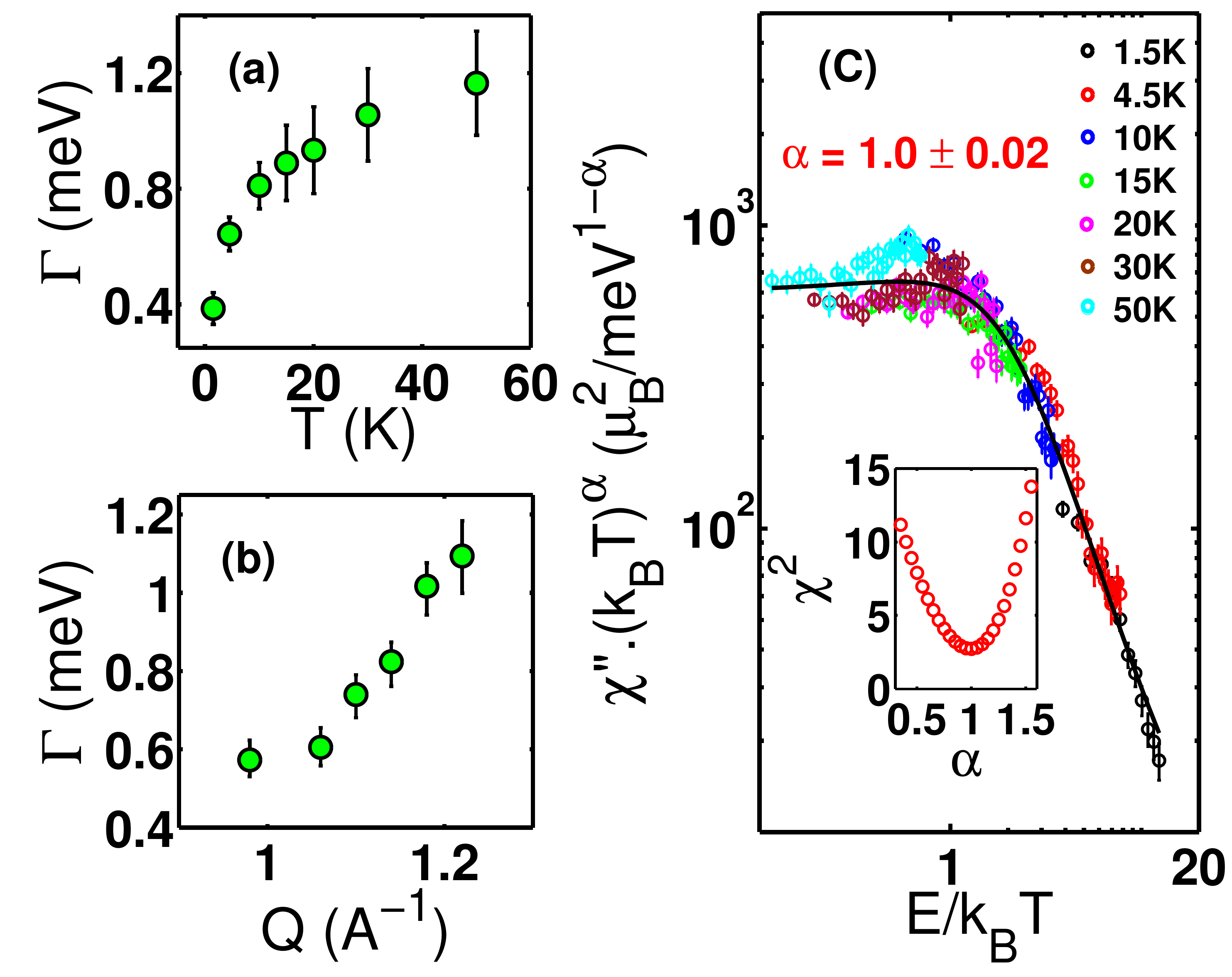} \vspace{-4mm}
\caption{(color online) (E/T) scaling and the quantum critical nature of spin fluctuations. (a) Estimated line-width $\Gamma$ ($\propto$ 1/$\tau$, $\tau$ - mean relaxation time) is plotted as a function of temperature at $Q$ = 1.06 $\AA$. The line-width decreases significantly below $T$$\simeq$ 25 K and approaches the instrument resolution ($\simeq$ 0.16 meV) as $T$$\rightarrow$~0 K, suggesting the presence of quantum critical behavior in the system. (b) $\Gamma$ as a function of absolute $Q$ at $T$ = 4.5 K. The line-width increases as a function of increasing Q. (c) (E/T) scaling of the dynamic susceptibilities further confirm the role of temperature as the single most relevant parameter of the spin fluctuations, as expected for the cooperative paramagnetic phenomena. Scaling was tried for various different values of the exponent $\alpha$. However, the best fit, given by minimum $\chi$$^{2}$, is obtained for $\alpha$ = 1.0 (see inset).
} \vspace{-4mm}
\end{figure}

Next, we investigate the quantum mechanical nature of the cooperative spin fluctuations in CaRuO$_{3}$. As inferred from Fig. 3e-h, inelastic peaks become broader as the measurement temperature is increased. The quantitative estimation of the line-width $\Gamma$ due to the quasi-elastic fluctuations of Ru$^{4+}$-ions, related to $\Gamma_1$ via $\Gamma$ = 2$\Gamma$$_{1}$$\surd$($\surd$2-1), suggests a significant slowing down of the mean relaxation time ($\tau$$\propto$~1/$\Gamma$) below $T$$\simeq$~25 K. $\Gamma$ tends to approach the instrument resolution limit as $T$$\rightarrow$0 K (Fig. 4a), indicating the development of a quantum critical state in the system.\cite{Singh}The estimated $\Gamma$ is atleast three times broader than the instrument resolution at different $Q$-values, Fig. 4b. More insight into the quantum-mechanical nature of the spin fluctuations is obtained by performing the scaling analysis of the dynamic spin susceptibility data in the form of $\chi$$^{"}$$T$$^{\alpha}$=$f(E/T)$, where $\alpha$ is the dynamic scaling exponent.\cite{Continentino} The dynamic susceptibilities are extracted from the inelastic neutron data using standard neutron scattering intensity analysis, given by $I$ = (1/$\pi$)$\chi$$^{"}$/(1-exp$^{-E/KT}$).\cite{Singh} The scaling plot of the dynamic susceptibilities is shown in Fig. 4c. As illustrated in the inset of Fig. 4c, the best overlap of the dynamic susceptibilities at different temperatures is obtained for the exponent $\alpha$ = 1(0.02). There are two important implications of this analysis: first, the linear ($E/T$) scaling reflects a Curie-Weiss type fluctuations, which is consistent with our argument that the cooperative paramagnetism is at the core of the observed spin fluctuations with temperature as the most relevant parameter.\cite{Schroder,Si} Second, it suggests the existence of a quantum critical fixed point in CaRuO$_{3}$. While more research is needed, preferably on single crystal specimen, to further understand the nature of the quantum critical fluctuations and the anisotropic properties, the cooperative nature of the paramagnetic fluctuations of Ru$^{4+}$ domains makes it a $^{'}$quasi-local$^{'}$ event.

We have performed detailed electrical, magnetic and neutron scattering measurements on the high purity powder sample of CaRuO$_{3}$ to investigate the spin dynamics and its correlation to the non-Fermi liquid behavior. It is found that Ru$^{4+}$ spins, residing in field-independent domains, exhibit strong paramagnetic fluctuations below $T$$\simeq$~ 22 K, which is very close to the cross-over temperature separating two NFL phenomena at low temperature. The field-independent characteristic of magnetic domains can also be arising due to a copling between the Ru$^{4+}$ spins fluctuations and the orbital fluctuations. A recent report has suggested that the orbital fluctuations play key role in the anomalous non-Fermi liquid properties in CaRuO$_{3}$.\cite{Laad} Further theoretical investigations are necessary to understand this. Nonetheless, the cooperative nature of the paramagnetic fluctuations lead to a dynamic ground state in CaRuO$_{3}$. As $T$$\rightarrow$0 K, the mean relaxation time of the paramagnetic fluctuations tends to become infinity. The divergence of the mean relxation time in conjunction with the linear ($E/T$) scaling of the dynamic susceptibilities hints of a quasi-local critical behavior, co-existing with the non-Fermi liquid state in this compound. The new findings provide a new research arena to explore new quantum material where the cooperative paramagnetism leads to a fragile or non-magnetic ground state, such as a quantum spin liquid.\cite{Mirebeau}

We acknowledge support from the University of Missouri Research Board and IGERT research program, funded by NSF under grant number DGE-1069091.

\clearpage

\end{document}